\documentclass[11pt,a4paper]{article}
\pdfoutput=1

\usepackage{jcappub}
\usepackage{graphicx}
\usepackage{hyperref}
\usepackage{amsmath}
\usepackage{amssymb}
\usepackage{color}

\providecommand{\f}[2]{\frac{{#1}}{{#2}}}
\newcommand{\be}{\begin{equation}} 
\newcommand{\ee}{\end{equation}}
\newcommand{\bea}{\begin{equation}\begin{aligned}} 
\newcommand{\eea}{\end{aligned}\end{equation}}

\newcommand{\td}{{\rm d}}

\def\lsim{\mathrel{\raise.3ex\hbox{$<$\kern-.75em\lower1ex\hbox{$\sim$}}}}
\def\gsim{\mathrel{\raise.3ex\hbox{$>$\kern-.75em\lower1ex\hbox{$\sim$}}}}

\begin{document}

\mbox{} \hfill IMPERIAL/TP/2019/TM/01 \\
\mbox{} \hfill KCL-PH-TH/2019-33

\title{Primordial Black Holes from \\Thermal Inflation}

\author[a]{Konstantinos Dimopoulos,}
\author[b,c]{Tommi Markkanen,}
\author[c]{Antonio Racioppi}
\author[d]{and Ville Vaskonen}

\affiliation[a]{Consortium for Fundamental Physics, Physics Department, Lancaster University, Lancaster LA1 4YB, UK}  
\affiliation[b]{Department of Physics, Imperial College London, Blackett Laboratory, London, SW7 2AZ, UK}
\affiliation[c]{National Institute of Chemical Physics and Biophysics, R\"{a}vala 10, 10143 Tallinn, Estonia}
\affiliation[d]{Physics Department, King's College London, London WC2R 2LS, UK}
                            
\emailAdd{k.dimopoulos1@lancaster.ac.uk}
\emailAdd{t.markkanen@imperial.ac.uk}
\emailAdd{tommi.markkanen@kbfi.ee}
\emailAdd{antonio.racioppi@kbfi.ee}
\emailAdd{ville.vaskonen@kcl.ac.uk}

\abstract{We present a novel mechanism for the production of primordial black holes (PBHs). The mechanism is based on a period of thermal inflation followed by fast-roll inflation due to tachyonic mass of order the Hubble scale. Large perturbations are generated at the end of the thermal inflation as the thermal inflaton potential turns from convex to concave. These perturbations can lead to copious production of PBHs when the relevant scales re-enter horizon. We show that such PBHs can naturally account for the observed dark matter in the Universe when the mass of the thermal inflaton is about $10^6\,$GeV and its coupling to the thermal bath preexisting the late inflation is of order unity. We consider also the possibility of forming the seeds of the supermassive black holes. {In this case we find that the mass of the thermal inflaton is about $1\,$GeV, but its couplings have to be very small, $\sim 10^{-7}$.} Finally we study a concrete realisation of our mechanism through a running mass model.}

\maketitle

\section{Introduction}
\label{introduction}
After their detection by LIGO~\cite{Abbott:2016blz}, black holes (BHs) have been the object of growing interest. In cosmology, there are various opportunities to form BHs in the early Universe, in which case they are called primordial BHs (PBHs)~\cite{Carr:1974nx}. Such PBHs can be much different from those formed from the collapse of stars. Indeed, they may be much lighter, which could lead to their fast evaporation~\cite{Hawking:1974rv}. If they form sufficiently heavy, $M\gsim 4\times 10^{14}\,$g, they have not evaporated until today and can form all, or a significant fraction, of the dark matter in the Universe. On the other hand, compared to astrophysical BHs, PBHs can already initially be much heavier and, since they form much earlier, they have more time to grow via accretion. They are therefore excellent candidates for the seeds of the supermassive BHs (SMBHs)~\cite{Bean:2002kx,Kawasaki:2012kn,Carr:2018rid}, whose origin remains elusive. Moreover, the BHs observed by LIGO and Virgo are relatively heavy and their spins are low~\cite{LIGOScientific:2018mvr}, which makes the astrophysical explanation for their origin challenging. PBHs, instead, naturally have very low spins~\cite{DeLuca:2019buf,Mirbabayi:2019uph}, and their merger rate has been shown to be in agreement with the LIGO and Virgo observations~\cite{Sasaki:2016jop}. These aspects have recently motivated various works considering PBH formation and constraints.

Inflation is generally accepted as the mechanism for the generation of the curvature perturbation in the Universe, which is reflected in the CMB primordial anisotropy and provides the seeds for structure formation. Many mechanisms which lead to PBH formation consider the possibility of a spiked curvature perturbation spectrum due to inflation provided there is a corresponding feature in the inflationary potential~\cite{Kawasaki:1997ju,Drees:2011hb,Drees:2011yz,Lyth:2011kj,Kawasaki:2012wr,Clesse:2015wea,Garcia-Bellido:2017mdw,Ezquiaga:2017fvi,Kannike:2017bxn}. In this paper we consider a new possibility to create PBHs. Instead of relying on the production of large perturbations during the primordial inflation, we study the generation of PBHs due to a late-time inflation period caused by thermal effects on the scalar potential. Such inflation, called thermal inflation, has been originally introduced in order to dilute dangerous relics such as moduli fields, gravitinos or topological defects formed at the end of primordial inflation~\cite{Lyth:1995hj,Lyth:1995ka}. 

Thermal inflation ends with a phase transition, which, as we show, can lead naturally to copious PBH production. After the thermal inflation, the field rolls down the potential towards its minimum, leading to a period of fast-roll inflation~\cite{Linde:2001ae}, which requires that the tachyonic mass\footnote{By tachyonic mass we mean that $m^2<0$.} of the inflaton field is comparable to the Hubble scale $m\sim H$. This is natural in supergravity theories~\cite{Dine:1995uk,Dine:1995kz}. We do not consider the case of a slow-rolling field with $m\ll H$ because this results in slow-roll after the end of thermal inflation, which takes a very large number of $e$-folds to complete. Consequently, the cosmological scales would leave the horizon during this slow-roll phase, rendering the phase transition and the thermal inflation phase irrelevant because they would correspond to superhorizon scales today. Moreover, we would need to produce the observed curvature perturbation that seeds structure formation during the slow-roll inflation. Even if the vacuum expectation value (VEV) of the thermal inflaton were such that slow-roll inflation would not result in a large number of $e$-folds, the possibility of $m\ll H$ also suffers from the danger of eternal inflation just before the onset of the rolling phase. Indeed, the slope of the potential near the hilltop is \mbox{$|V'|=m^2\phi$}, where $m$ is the tachyonic mass of the field $\phi$. Near the onset of the phase transition, the field is displaced from the minimum by quantum fluctuations as \mbox{$\phi^2\sim H^2$}, which means that if \mbox{$m\ll H$}, we have \mbox{$|V'|\ll H^3$}. This suggests that the variation of the field is dominated by its quantum fluctuations, which leads to eternal inflation because there would always be locations where the quantum fluctuations conspire to keep the field on top of the potential hill. Even if not eternal, inflation could continue in some locations much more than in others, which is unacceptable on subhorizon scales because it leads to excessive curvature perturbations that would be in conflict with observations.

The mechanism of thermal inflation is rather general and does not point towards a specific model. In this context we calculate the production of PBHs and show that they can account for the total of dark matter if the bare mass of the thermal inflaton is $m\sim 10^6\,$GeV, or they can provide the seeds for the SMBHs if $m\sim 1\,$GeV. In an effort to solidify our mechanism, we also discuss a concrete realisation by considering a running mass model due to loop corrections. 

Disentangling PBH formation from primordial inflation has the merits that our mechanism is generic and does not require a specific model of primordial inflation. However, the presence of a late inflation period (thermal plus fast-roll inflation) does alter inflationary predictions remarkably. As a result, the usual plateau models of inflation, like Starobinsky~\cite{Starobinsky:1980te} or Higgs inflation~\cite{Bezrukov:2007ep}, do not agree with the observations anymore. However, other equally well motivated models of primordial inflation can be employed. In this paper we discuss minimal hybrid inflation in supergravity~\cite{Linde:1997sj,Dvali:1997uq} as an example of a primordial inflation model that has recently in Ref.~\cite{Dimopoulos:2016tzn} been shown to be successful.

The paper is organized as follows: First, in Sec.~\ref{sec:Largepertubations} we introduce a generic model for thermal inflation and calculate the curvature power spectrum produced during the late inflation. In Sec.~\ref{sec:PBH} we then calculate the spectrum of PBHs produced from those perturbations, and show what parameters are needed for them to provide the dark matter or the SMBH seeds. We consider the effects of the late inflation for the parameters of interest on the predictions of the primordial inflation in Sec.~\ref{sec:primeinf}, and finally in Sec.~\ref{sec:runningmass} we introduce an example of how the thermal inflaton potential could be stabilised. Our main conclusions are summarised in~Sec.~\ref{sec:conclusions}.

We use natural units where $c=\hbar=1$, and the reduced Planck mass $m_P=2.43\times 10^{18}\,$GeV. Our sign conventions are $(+,+,+)$ in the classification of~\cite{Misner:1974qy}.

\section{Perturbations from thermal inflation}
\label{sec:Largepertubations}
Let us postulate a generic finite temperature potential for a scalar field with a phase transition and non-zero vacuum energy $V_0$ at the origin,
\be
V(\phi) = V_0-\frac12\left(m^2-g^2T^2\right) \phi^2 + \dots\,,\label{eq:pot}
\ee
where $m>0$ is a mass parameter inducing the onset of the phase transition at $gT = m$. The ellipsis denote at the moment unspecified stabilising terms allowing a vacuum expectation value of $\phi$ to be of the order $\phi_{\rm vev}\sim m_P$ and $V_0>0$ is a constant density scale necessary to avoid negative vacuum density in the broken phase. The parameter $g>0$ is a coupling of $\phi$ to a pre-existing thermal bath of temperature $T$. Starting from a large temperature, $T\gg m/g$, the thermal correction stabilises the potential by introducing a positive contribution to the effective mass,
\be
m_{\rm eff}^2\equiv -m^2+g^2T^2\,. \label{eq:meff}
\ee
By setting $a=1$ at $T=m/g$ and since the temperature dilutes as $\propto a^{-1}$, we can write $g T = m/a$ and the effective mass becomes
\be \label{eq:effm}
m_{\rm eff}^2 = -m^2(1-a^{-2})\,,
\ee
normalizing the onset of the phase transition to take place at $a=1$.

{From the point of view of PBH creation thermal inflation possesses the attractive feature that during the phase transition the potential goes from convex to concave, which is expected to (and as we show will) lead to a spike in curvature perturbations. It is hence not necessary to tune the potential to have a specific feature or a very flat plateau, in contrast to what is often required for PBH generation.}

Assuming initially the system to be in thermal equilibrium after sufficient dilution the vacuum energy $V_0$ starts to dominate the energy density leading 
a period of (thermal) inflation~\cite{Lyth:1995ka}. This takes place when the radiation energy density drops below $V_0$, that is, at
\be
T_{\rm beg} = \left(\frac{30}{\pi^2 g_*}\right)^{1/4} V_0^{1/4} \simeq 0.4 V_0^{1/4}\,.
\ee
In the last step we assumed that the effective number of relativistic degrees of freedom at $T_{\rm beg}$ is $g_*\simeq 100$ and the Hubble rate is then given by $H^2 = V_0/3m_P^2$. The number of $e$-folds before the onset of the phase transition at $T_{\rm end}=m/g$ is then
\be \label{eq:NT}
N_T = \ln\left(\frac{T_{\rm beg}}{T_{\rm end}}\right) \simeq \ln\left(\frac{0.4 V_0^{1/4}}{m}\right) + \ln g \,.
\ee

Let us next consider the behaviour of the quantum modes and the curvature power spectrum. Since the potential goes from convex to concave with a tachyonic mass, one expects non-trivial dynamics for the quantum evolution and a spike in the generated curvature perturbation. Fortunately, for a quadratic theory as given by Eq.~\eqref{eq:pot} the evolution of the quantum fields may be solved without perturbation theory when the background is taken to be fixed.%
\footnote{When calculating the evolution of the quantum modes we take the background to be strictly de~Sitter. This is a very good approximation close to the onset of the phase transition, since our focus is on the parameter range $m\sim H$ and, since the field starts in its vacuum, there is (crudely) only one scale relevant for the change of the energy density during the phase transition, $|\dot{\rho}|\sim H^5$ ($|\dot\rho|\sim\Delta\rho/\Delta t$ with $\Delta\rho\sim H^4$ per Hubble time $\Delta t\sim 1/H$), giving via the Friedmann equation $\rho=3H^2m_P^2$
$$
|\epsilon|=\frac{|\dot{H}|}{H^2}\sim\frac{|\dot{\rho}|}{H^3 m_P^2}\sim \frac{H^2}{m_P^2}\ll1\,.
$$
}

In conformal time in de Sitter space 
\be\eta=-(a H)^{-1}\,;\quad a=e^{Ht}\,,
\ee
the properly normalised mode is given by
\be \label{eq:adsol2}
\hat{\phi}=\int \frac{\td^{3}{k}}{\sqrt{(2\pi )^{3}a^2}}\left[\hat{a}_\mathbf{k}^{\phantom{\dagger}}u^{\phantom{\dagger}}_{k}(\eta)+\hat{a}_{-\mathbf{k}}^\dagger u^*_k(\eta)\right]e^{i\mathbf{k\cdot\mathbf{x}}}\,,
\ee
where $\mathbf{k}$ is the co-moving momentum, $k\equiv|\mathbf{k}|$ and the ladder operators are normalized as $[\hat{a}_{\mathbf{k}}^{\phantom{\dagger}},\hat{a}_{\mathbf{k}'}^\dagger]=\delta^{(3)}(\mathbf{k}-\mathbf{k}')$. The equation of motion $(-\Box+m^2_{\rm eff})\hat{\phi}=0$, is then
\be
{u}''_k(\eta)+\bigg(k^2+m^2 - \frac{\nu^2-1/4}{\eta^2}\bigg){u}_k(\eta)=0\,,
\ee
where 
\be \label{eq:nu}
\nu^2\equiv \frac{9}{4}+\frac{m^2}{H^2}\,.
\ee
The Bunch-Davies (BD) vacuum \cite{Bunch:1978yq,Chernikov:1968zm} corresponds to choosing
\be \label{eq:bd}
{u}_k(\eta)=\frac{1}{2}\sqrt{-\pi\eta}H^{(1)}_\nu \left(-\sqrt{k^2+m^2}\,\eta\right)\,,
\ee
with $H^{(1)}_\nu$ being the Hankel function of the first kind. Since the BD vacuum is an attractor state in de Sitter space~\cite{Anderson:2000wx,Finelli:2008zg,Markkanen:2016aes} for any mode exiting the horizon during inflation the BD vacuum is a well-motivated boundary condition. For our set-up this implies that calculating the curvature power spectrum in the BD vacuum is justified close to the onset of the phase transition if there has been several $e$-folds of inflation before the onset of the phase transition. This turns out to be satisfied in the parameter range we are interested in.

For our purposes only large scale fluctuations are relevant, so we can use the infra-red (IR) limit that is defined by including only the superhorizon modes for which we make use of the asymptotic expansion for the Hankel functions and the integral representation for the ordinary hypergeometric function~\cite{AbramowitzStegun64}
\be \label{eq:asymp0}
H^{(1)}_\alpha(x)\simeq-\frac{i 2^{\alpha } \Gamma (\alpha ) x^{-\alpha }}{\pi }\,;\qquad x\ll\sqrt{\alpha +1}\,,
\ee
\be
{\rm B}(b,c-b)\,_2F_1(a,b;c;z) = \int_0^1 dx\,
x^{b-1} (1-x)^{c-b-1}(1-zx)^{-a} \, \,;\quad \Re (c) > \Re(b) > 0\,; z<1\,,
\ee
where ${\rm B}(a,b)$ is the Euler Beta function, to find an expression for the variance
\bea \label{eq:var}
\langle\hat{\phi}^2\rangle &= \int_{\rm IR}\frac{\td^3k }{(2\pi )^{3}a^2} |u_k(\eta)|^2 = \int_{0}^{a H} \frac{\td k\,k^2 }{2\pi^{2}a^2} |u_k(\eta)|^2  \\ 
&= \left(\frac{H}{2\pi}\right)^2 \frac{2^{2\nu}\Gamma(\nu)^2}{6\pi} \left(\frac{a^2H^2}{m^2}\right)^{\nu} \mbox{}_2F_1\left[\frac{3}{2},\nu;\frac{5}{2};-\frac{a^2H^2}{m^2}\right]\,.
\eea
Setting the lower limit of the integral to 0 is justified again if the system has been inflating a sufficiently long time\footnote{Setting a lower cut off at the scale that exited the horizon at the start of thermal inflation this translates as demanding $e^{-N_T}\ll1$.}.

The late time limit of the variance can be calculated via
\bea
_2F_1 (b,a;c;z) &=\,_2F_1 (a,b;c;z) = (1-z)^{-b} {}_2F_1 \bigg [b,c-a;c;\tfrac{z}{z-1} \bigg]\\&\overset{z\rightarrow-\infty}{=}z^{-b} {}_2F_1 \bigg [b,c-a;c;1 \bigg]=z^{-b}\f{\Gamma(c)\Gamma(a-b)}{\Gamma(c-b)\Gamma(a)}\,;\quad \Re(a-b)>0\,, 
\eea
giving
\be \label{eq:as}
\langle\hat{\phi}^2\rangle\overset{t\rightarrow\infty}{\longrightarrow}\bigg(\f{H}{2\pi}\bigg)^2\frac{\Gamma \left(\nu -\frac{3}{2}\right) \Gamma (\nu ) }{\sqrt{\pi }}\bigg(\f{2aH}{m}\bigg)^{2\nu-3} \,,
\ee
which shows that the modes will grow without bound if no stabilising correction is introduced, as is of course expected. Note that this happens also at the massless limit, which is a manifestation of the well-known secular growth exhibited by a massless scalar spectator field in de Sitter space~\cite{Linde:1982uu,Vilenkin:1982wt,Starobinsky:1982ee,Allen:1985ux}.

Calculating the curvature perturbation proceeds in the usual manner and can be performed analytically by using the modes~\eqref{eq:bd}. However, it is important to note that
after the onset of the phase transition there is no classical inflaton rolling down a potential: The system is symmetric with a strictly vanishing vacuum expectation value (VEV), $\langle\hat{\phi}\rangle=0$. This does not mean that no perturbations are generated, but rather that the system is not linear and the spectrum must by calculated by using the full (non-linear) result.

The curvature perturbation in the spatially flat gauge can be written as
\be \label{eq:zeta}
{\mathcal{\zeta}}=\f{H}{\dot{\rho}}\delta{\rho}\approx H\f{V({\phi})-\langle V({\phi})\rangle}{\partial_t{\langle V}({\phi})\rangle}=\f{H m^2_{\rm eff}}{\partial_t[m^2_{\rm eff}\langle\hat{\phi^2}\rangle]}\delta\phi^2\,,
\ee
where the gradient terms were dropped from the energy-density ${\rho}$ as sub-leading and the fluctuation is defined as the difference to the mean $\delta {\rho} \equiv {\rho}-\langle{\rho}\rangle$. { Furthermore, we have neglected the energy density of the thermal bath, which we will justify shortly.} The power spectrum of a function $f$ is defined by the Fourier transform
\be \label{eq:P}
{\cal P}_f(k)=\frac{k^3}{2\pi^2}\int d^3x\, e^{i\mathbf{k}\cdot\mathbf{x}}\langle{f}(0){f}({\bf x})\rangle\,,
\ee
which with~\eqref{eq:zeta} gives
\be \label{eq:nonl}
{\cal P}_\zeta (k) = \bigg(\f{H m^2_{\rm eff}}{\partial_t[m^2_{\rm eff}\langle\hat{\phi^2}\rangle]}\bigg)^2{\cal P}_{\delta\phi^2} (k)\,.
\ee
The ${\cal P}_{\delta\phi^2} (k)$ is obtained from the quantum fields Eq.~\eqref{eq:adsol2} with Wick's theorem
\bea
{\cal P}_{\delta\phi^2} (k)
&=\f{k^3}{2\pi^2}\int d^3x\, e^{i \mathbf{k}\cdot\mathbf{x}}\big[\langle\delta \hat{\phi}^2(0)\delta \hat{\phi}^2(\mathbf{x})\rangle\big] \\
&=\f{k^3}{2\pi^2}\int d^3x\, e^{i \mathbf{k}\cdot\mathbf{x}}\big[\langle(\hat{\phi}^2(0)-\langle\hat{\phi}^2\rangle)(\hat{\phi}^2(\mathbf{x})-\langle\hat{\phi}^2)\rangle\big] \\
&=\f{k^3}{2\pi^2}\int d^3x\, e^{i \mathbf{k}\cdot\mathbf{x}}\big[\langle\hat{\phi}^2(0)\hat{\phi}^2(\mathbf{x})\rangle-\langle\hat{\phi}^2\rangle^2\big] \\
&=\f{k^3}{\pi^2}\int \f{d^{3}{p}}{{(2\pi )^{3}}}\f{1}{a^4} |u^{\phantom{\dagger}}_{p}(\eta)|^2|u^{\phantom{\dagger}}_{|\mathbf{k}-\mathbf{p}|}(\eta)|^2\,.
\eea
With the asymptotic form valid at the IR~\eqref{eq:asymp0} and aligning the $\mathbf{p}$-coordinates with $\mathbf{k}$ to perform the integrals over the polar and azimuthal angles we obtain 
\bea{\cal P}_{\delta\phi^2} (k)=&\f{2^{4\nu-7}\Gamma(\nu)^4}{\pi^6}\f{H^4}{1-\nu} \int_0^1 dz\,r^2z\left(z^2+y^2\right)^{-\nu} \\ &\times\bigg\{\bigg[(z+r)^2+y^2\bigg]^{1-\nu}-\bigg[(z-r)^2+y^2\bigg]^{1-\nu}\bigg\}\,,\label{eq:Pphi2}
\eea
with the definitions~\eqref{eq:nu} and
\be
r\equiv \f{k}{aH}\,,\quad y\equiv \f{m}{aH}\,.
\ee





For many cases, reasonable results are expected from the root-mean-square approximation for the mean field
\be \label{eq:fluk}
\varphi\equiv\sqrt{\langle \hat{\phi}^2\rangle}\,,
\ee
which can be used to write the power spectrum to linear order in fluctuations around Eq.~\eqref{eq:fluk} 
\be{\mathcal{\zeta}}=\f{H}{\dot{\rho}}\delta{\rho}\approx H\f{V'({\varphi})}{\partial_t[ V(\varphi)]}\delta{\phi}
\,.
\ee
With the help of~\eqref{eq:var} this gives the linear approximation to the power spectrum
\bea \label{Pzeta}
\mathcal{P}_\zeta(k) &\approx\bigg( H\f{V'({\varphi})}{\partial_t[ V(\varphi)]}\bigg)^2\mathcal{P}_{\delta\phi}(k) \\&= \frac{12H^2m^4_{\rm eff}\big(\frac{a^2H^2}{m^2}\big)^{\nu}\,_2F_1\big[\frac{3}{2},\nu;\frac{5}{2},-\frac{a^2H^2}{m^2}\big]}{\big\{\partial_t\big[m^2_{\rm eff}\big(\frac{a^2H^2}{m^2}\big)^{\nu}\,_2F_1\big[\frac{3}{2},\nu;\frac{5}{2},-\frac{a^2H^2}{m^2}\big]\big]\big\}^2} 
\bigg(\frac{k}{aH}\bigg)^3\bigg[\frac{k^2+m^2}{a^2H^2}\bigg]^{-\nu} \,.
\eea
In this approximation the spectrum peaks at 
\be 
k_{\rm max} = \frac{H}{2}\sqrt{3(2\nu + 3)} \,.
\ee

Although the linear approximation in Eq.~\eqref{Pzeta} gives results that are in good agreement with the non-linear expression in Eq.~\eqref{eq:nonl}, when calculating the PBH yield this may lead to inaccuracies due to its exponential dependence on  the power spectrum (see Eq.~\eqref{eq:PS}). For this reason we will only make use of Eq.~\eqref{eq:nonl} for the results presented in Sec.~\ref{sec:PBH}.

\begin{figure}
\centering
\includegraphics[height=0.32\textwidth]{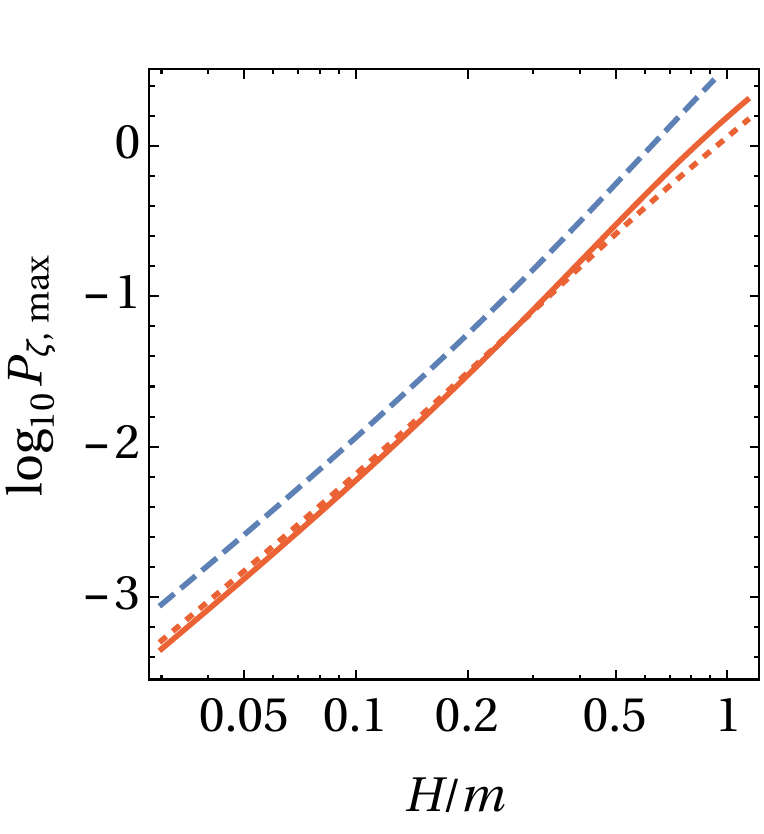} \hspace{4mm}
\includegraphics[height=0.32\textwidth]{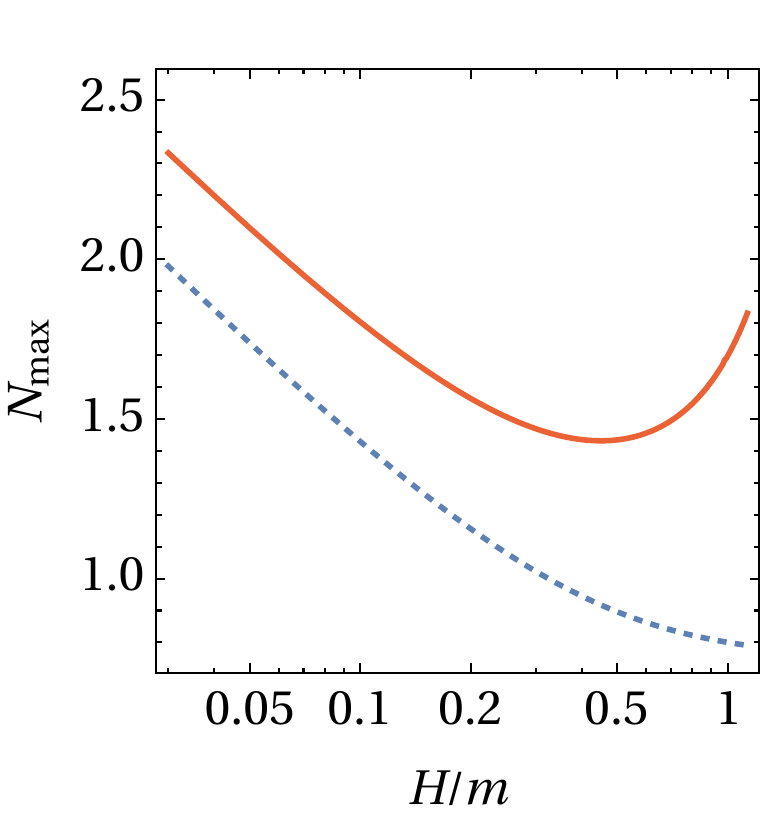} \hspace{4mm}
\includegraphics[height=0.32\textwidth]{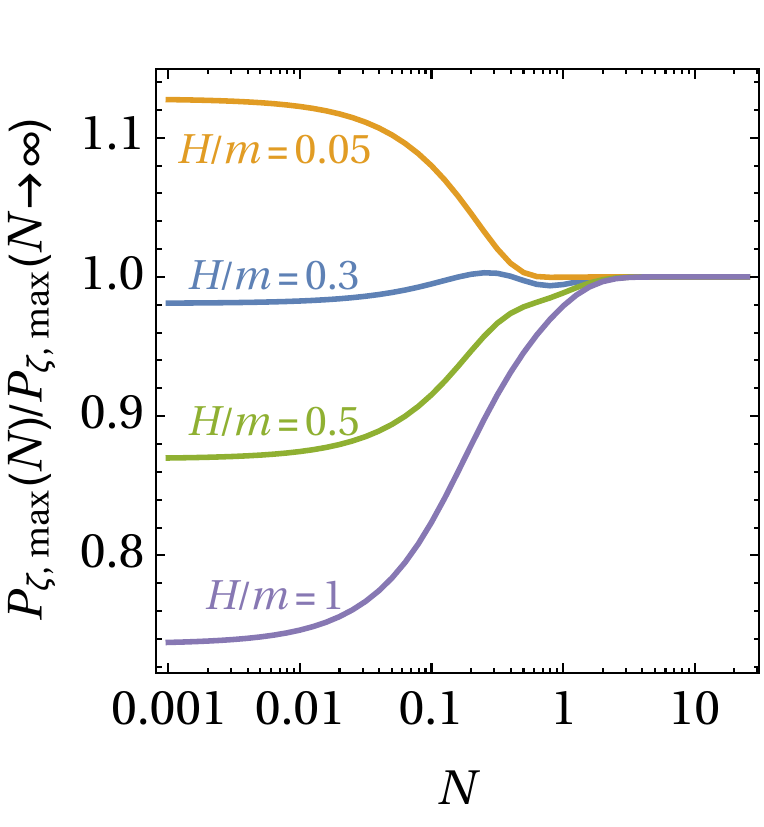}
\caption{The peak amplitude of the curvature power spectrum $P_{\zeta,{\rm max}} = P_\zeta(k_{\rm max})$ (left panel) and the number of $e$-folds $N_{\rm max}$ when the scale $k_{\rm max}$ exits horizon after $a=1$ (middle panel) as a function of $H/m$. The blue dashed lines correspond to the approximation~\eqref{Pzeta}, and the red ones to the full result~\eqref{eq:Pphi2}. In the left panel the dotted red line shows the amplitude at the horizon exit of the scale $k_{\rm max}$ and the solid one the asymptotic limit $N\to \infty$. The right panel shows the evolution of the peak amplitude as a function of the number of e-folds after $a=1$.}
\label{fig:Pmax}
\end{figure}

The curvature power spectrum in Eq.~\eqref{eq:nonl} depends only on the mass parameter $m$ and the Hubble rate during inflation, $H^2 = V_0/3m_P^2$, but not on $g$. By numerically evaluating the integral in Eq.~\eqref{eq:Pphi2}, we find that the peak amplitude of $\mathcal{P}_\zeta(k)$ is a function of $H/m$ only. This dependence is shown in the left panel of Fig.~\ref{fig:Pmax}. Consistently with the approximation~\eqref{Pzeta}, the position of the peak, $k=k_{\rm max}$, is directly proportional to m, $k_{\rm max}\propto m$. In the middle panel of Fig.~\ref{fig:Pmax} we show when the scale $k_{\rm max}$ exits horizon, measured in $e$-folds from $a=1$. The right panel shows instead that outside horizon the spectrum becomes independent of $a$. At scales smaller than $k_{\rm max}$ the spectrum increases as $k^3$, and at larger scales it decreases as $k^{3-2\nu}$, as is also borne out by the linear approximation~\eqref{Pzeta}. 

\begin{figure}
\centering
\includegraphics[height=0.36\textwidth]{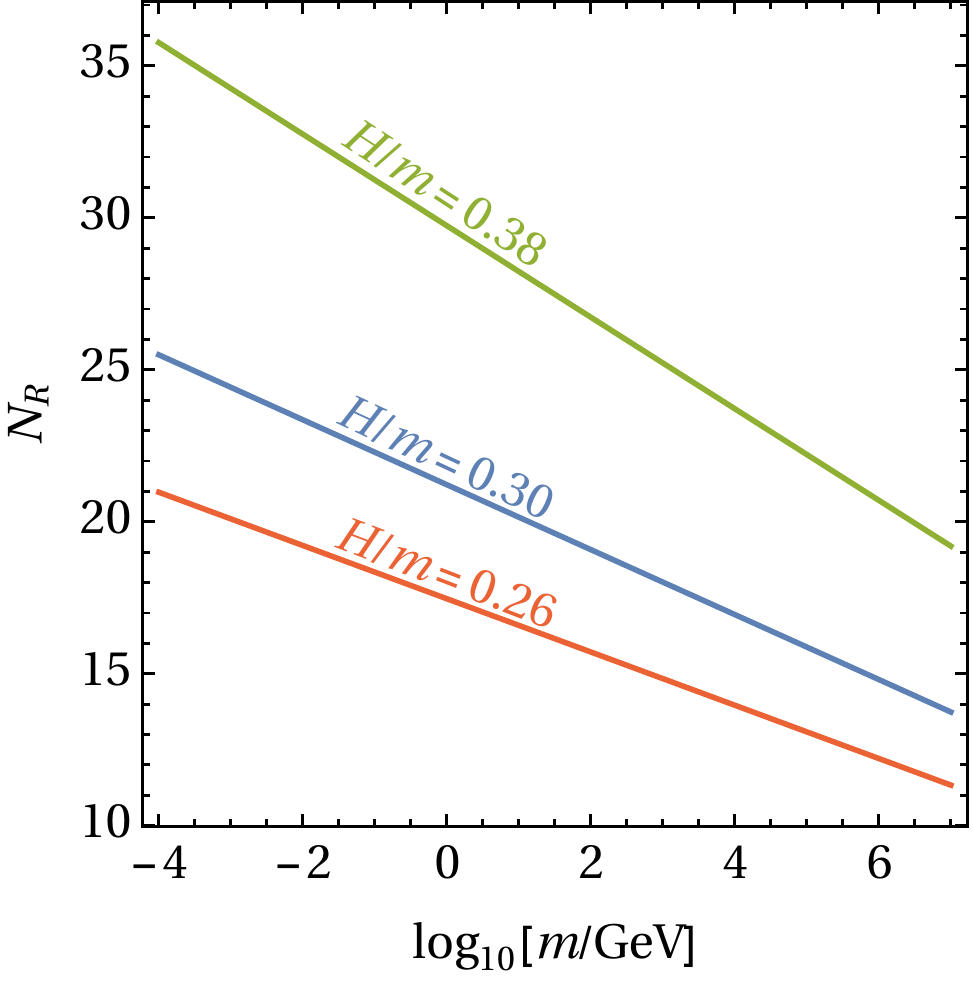} \hspace{10mm}
\includegraphics[height=0.36\textwidth]{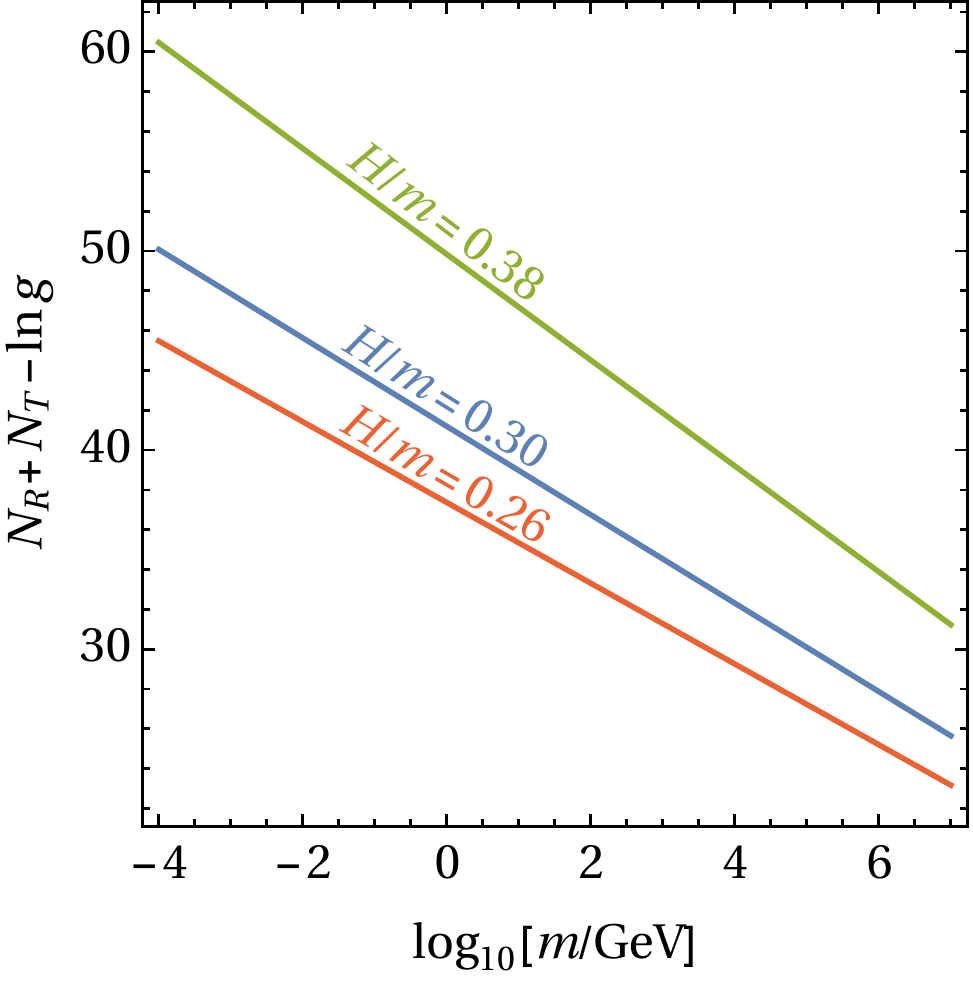} 
\caption{The number of $e$-folds of the thermal inflation and the fast-roll inflation as a function of the mass parameter $m$ for three values of $H/m$.}
\label{fig:NR}
\end{figure}

To calculate the length of the inflationary period after $T_{\rm end}=m/g$ we use the root-mean-squared (RMS) field~\eqref{eq:fluk}. The inflation ends when $\dot\varphi^2 \simeq 2V(\varphi)$ as the kinetic energy starts to dominate over the vacuum energy. This happens only slightly before the minimum of the potential is reached, $\varphi^2 \sim V_0/m^2$, and we approximate the end of inflation by the latter.\footnote{We note that most of the $e$-folds arise when the field is close to $\phi=0$. So, the approximation of determining the end of inflation by finding when $\varphi^2 \sim V_0/m^2$ is reliable independently of the mechanism that stabilises the potential.} The number of $e$-folds of inflation at $T<m/g$ is then given by
\be \label{eq:NR}
N_R \simeq \ln\left( \frac{m}{H} \right) + \frac{1}{2\nu-3}\ln\left[ \frac{3\times 4^{3-\nu} \pi^{5/2} m_P^2}{m^2 \Gamma(\nu-3/2) \Gamma(\nu)} \right] \,,
\ee
where we used the late time asymptotic form in Eq.~\eqref{eq:as} for $\varphi$. The above agrees nicely with Ref.~\cite{Linde:2001ae}. In total, the inflationary period lasts $N_T + N_R$ $e$-folds. We have plotted $N_R$ and $N_R+N_T-\ln g$ as a function of $m$ for three values of $H/m$ in Fig.~\ref{fig:NR}.

{When calculating the curvature perturbation in Eq. (\ref{eq:zeta}) we have neglected the energy density of the thermal bath even though it may be non-negligible close to the phase transition. This can be motivated by realising that by the end of the fast roll phase the energy density of the thermal bath, $\sim V_0 e^{-4(N_T+N_R)}$, will be completely dominated by the energy density of the rolling field $\sim V_0$. Therefore, to a very good approximation the curvature perturbation we are left with after inflation will result entirely from the vacuum fluctuations of the scalar field, which is a separately conserved contribution as visible in Fig.~\ref{fig:Pmax}.
}

Afterwards, the $\phi$ field oscillates around the minimum of the potential, and decays into radiation reheating the plasma to temperature $T_{\rm reh}$. We point out that the products of the decay of the thermal inflaton field $\phi$ at the end of fast-roll inflation cannot be the same with the particles comprising the thermal bath which preexisted thermal inflation. This is because the field expectation value has changed by $\phi_{\rm vev}\sim m_P$, which means that after the end of fast-roll inflation the particles with which $\phi$ interacted via the coupling $g$ when kept at the origin by the thermal correction, have obtained masses that can be very large unless $g$ is tiny. As a result, $\phi$ can not decay into them if $m_V<gm_P$, where $m_V\sim m$ is the mass of the thermal inflaton particles in the vacuum. However, it would decay into other particles via another coupling $\hat g$, especially if the VEV is an enhanced symmetry point characterised by couplings of the form $\hat g(\phi-\phi_{\rm vev})\psi\bar\psi$, with $\psi$ being a fermion field.

If $\hat g$ is sufficiently large for $\phi$ to decay promptly at the end of the late inflation then
\be \label{eq:Trehprompt}
T_{\rm reh} \simeq T_{\rm beg} \simeq 0.4 V_0^{1/4} \,.
\ee
However, if $\hat g\ll 1$ then the decay rate of the thermal inflaton $\phi$, that we approximate as \mbox{$\Gamma\sim\hat g^2m$}, can be much smaller than the Hubble scale during the late inflation,  $H^2=V_0/3m_P^2$. The Universe then experiences a period of matter domination, as the thermal inflaton oscillates around the minimum of the potential, until the Hubble rate has decreased below $\Gamma$ and the thermal inflaton decays into
the radiation bath of the Hot Big Bang, thereby reheating the Universe to
\be \label{eq:Trehlate}
T_{\rm reh}\simeq 0.4(\Gamma/H)^{1/2}V_0^{1/4} \,.
\ee

\begin{figure}
\centering
\includegraphics[height=0.38\textwidth]{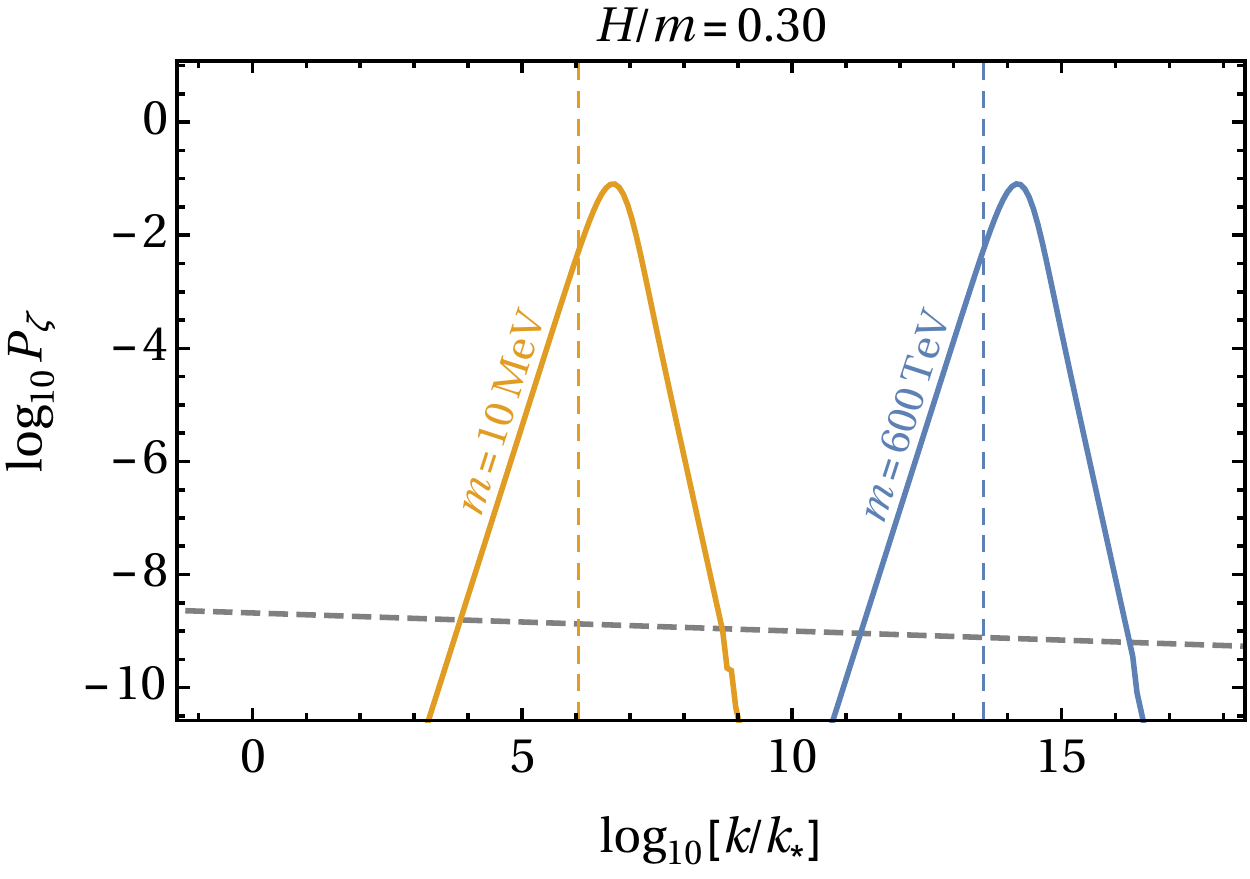} 
\caption{The solid lines show the curvature power spectrum produced during the thermal inflation. The vertical dashed lines highlight the scale that exits the horizon at $T_{\rm end}=m/g$. The gray dashed line shows extrapolation of the CMB observations.}
\label{fig:P}
\end{figure}

Finally, the curvature power spectrum for different values of the mass parameter $m$ is shown in Fig.~\ref{fig:P}. We have normalized the horizontal axis to the Planck pivot scale~\cite{Aghanim:2018eyx}, that in the usual normalization of the scale factor, where $a=1$ today, is $0.05\, {\rm Mpc}^{-1}$. In our normalisation the scale factor today is 
\be
a(T_0) = e^{N_R} \left[\frac{s(T_{\rm reh})}{s(T_0)}\right]^{1/3} \,,
\ee
where $s$ is the entropy density and $T_0\approx 2.35\times 10^{-13}\,$GeV is CMB temperature today. So, the Planck pivot scale is
\be
k_* \equiv 0.05\,{\rm Mpc}^{-1}\, a(T_0) \simeq 2.3\times 10^{-27} \,e^{N_R} \sqrt{H m_P} \,,
\ee
where we assumed prompt reheating and that the number of relativistic entropy degrees of freedom at $T_{\rm reh}$ is $\simeq 100$. The gray dashed line corresponds to $\mathcal{P}_\zeta(k) = A_s (k/k_*)^{n_s-1}$, where $A_s=2.14\times 10^{-9}$ is the observed amplitude of the curvature power spectrum at $k=k_*$ and $n_s=0.965$ is the scalar spectral index~\cite{Aghanim:2018eyx}. Especially at scales that exited horizon much before $a=1$ the assumption of BD vacuum may not be valid, and the spectra shown in Fig.~\ref{fig:P} may therefore overestimate the amplitude at low $k$. We are, however, interested only on the scales close to the peak of the spectrum at $k = k_{\rm max}$. For $H=0.30m$ the scale $k_{\rm max}$ exits horizon $1.5$ $e$-folds after $a=1$, and we have $3-2\nu=-4.3$, so the spectrum drops very fast above $k_{\rm max}$.

\section{Primordial black holes}
\label{sec:PBH}
As shown in the previous section, potentially large perturbations are produced around the time when the potential is flat at $\phi=0$, that is, $N_R$ $e$-folds before the end of inflation. Let us now consider a perturbation with a density contrast $\delta$ at a comoving scale $R$. If $\delta$ is larger than a threshold value $\delta_c$~\cite{Carr:1975qj}, part of the horizon mass collapses to a BH almost immediately when the scale $R$ re-enters horizon, $R = (a_R H_R)^{-1}$. This happens at 
\be \label{eq:aR}
a_R= \xi^{-\frac13} e^{2N_R} R H \,,
\ee
where $\xi \equiv \min(1,\Gamma/H)$. The mass of the produced PBH follows the critical scaling~\cite{Choptuik:1992jv,Niemeyer:1997mt,Niemeyer:1999ak},
\be \label{eq:Mcrit}
M_{{\rm BH},R}(\delta) = \kappa (\delta-\delta_c)^\gamma M_R \,,
\ee
where $M_R$ is the horizon mass at the horizon re-entry, and $\kappa$ and $\gamma$ are constants. We use $\delta_c \simeq 0.42$~\cite{Harada:2015yda} (see also Refs.~\cite{Shibata:1999zs,Musco:2004ak}), $k=3.3$~\cite{Evans:1994pj} and $\gamma=0.36$~\cite{Koike:1995jm}. The horizon mass is given by
\be
M_R = \rho_{{\rm tot},R} \f{4\pi}{3}H_R^{-3} = 4\pi a_R R m_P^2 \,,
\ee 
where $\rho_{{\rm tot},R}$ and $H_R$ are the total energy density and the Hubble rate at $a=a_R$. 

In order to calculate the fraction of the Universe content that collapses into BHs, we first relate the curvature perturbation power spectrum to the smoothed density contrast at the comoving scale $R$ (see e.g. Ref.~\cite{Young:2014ana}),
\be
\sigma_R^2 = \left(\frac{4}{9}\right)^2\int_0^\infty \frac{\td k}{k} e^{-k^2R^2} (k R)^4 \mathcal{P}_\zeta(k) \,.
\ee
In the late inflation model discussed in the previous section the generated fluctuations are purely Gaussian {up to quantum gravity corrections suppressed by $H^2/m_P^2$.} Hence we make use of
\be
P_R(\delta) =  \frac{1}{\sqrt{2\pi} \sigma_R} \exp\left( -\frac{\delta^2}{2\sigma_R^2} \right) \,.
\ee 
Assuming that the scale $R$ re-enters horizon during radiation domination the fraction of the universe that collapses into BHs at $a=a_R$,
\be \label{eq:beta}
\beta_R \equiv\frac{1}{\rho_{\gamma}(T_R)} \frac{\td\rho_R}{\td\ln R} \,,
\ee
is in the Press-Schechter approach,\footnote{{This approach has recently been questioned in Refs.~\cite{Yoo:2018esr,Germani:2018jgr} which suggest that a smaller value for $\sigma_R$ is needed. We, however, use the conventional Press-Schechter approach for its simplicity. We believe that the possible corrections will not change our results significantly.}} accounting for the critical collapse, given by~\cite{Niemeyer:1997mt}
\be \label{eq:PS}
\beta_R = \int_{\delta_c}^\infty \td\delta\, \frac{M_{{\rm BH},R}(\delta)}{M_R} P_R(\delta) \approx \kappa \sigma_R^{2\delta}\, {\rm Erfc}\left(\frac{\delta_c}{\sqrt{2} \,\sigma_R}\right) \,.
\ee
The last approximation holds for $\sigma\ll \delta_c$, which is valid in the case that {PBHs are not overproduced}. If the scale $R$ re-enters horizon during matter dominance the PBH production would get significantly enhanced~\cite{Khlopov:1980mg,Polnarev:1986bi,Carr:2018nkm}. We will not consider this possibility. We assume that the reheating process is fast enough so that the relevant scales re-enter horizon after that. We check that this assumption is valid for the parameters that we use by comparing the temperature when $k_{\rm max}$ re-enters horizon to the reheating temperature.

From the definition in Eq.~\eqref{eq:beta} we see that the energy density in the PBHs formed when the {scales in the logarithmic range $(\ln R, \ln R + \td \ln R)$ re-enter horizon is $\td \rho_R = \rho_\gamma(T_R) \beta_R \td\ln R$.} Neglecting accretion and mergers, and assuming the standard expansion history after the formation of the PBHs, their present abundance can be obtained by scaling this energy density as matter until today,
\be
\td \Omega_R = \frac{s(T_0)}{s(T_R)} \frac{\rho_\gamma(T_R)}{\rho_c} \beta_R \td\ln R \simeq 4.6\times 10^8 \frac{T_R}{\rm GeV} \beta_R \td\ln R\,,
\ee
where $\rho_c$ is the critical density, $T_0$ is the CMB temperature today. The temperature $T_R$ is obtained from entropy conservation,
\be
a_R^3 s(T_R) = \xi^{-2} e^{3N_R} s(T_{\rm reh}) \,,
\ee
and for the last step we approximated that at $T_R$ the effective numbers of relativistic entropy and energy degrees of freedom are equal. The total abundance of PBHs today is $\Omega = \int \td \Omega_R$.

\begin{figure}
\centering
\includegraphics[height=0.38\textwidth]{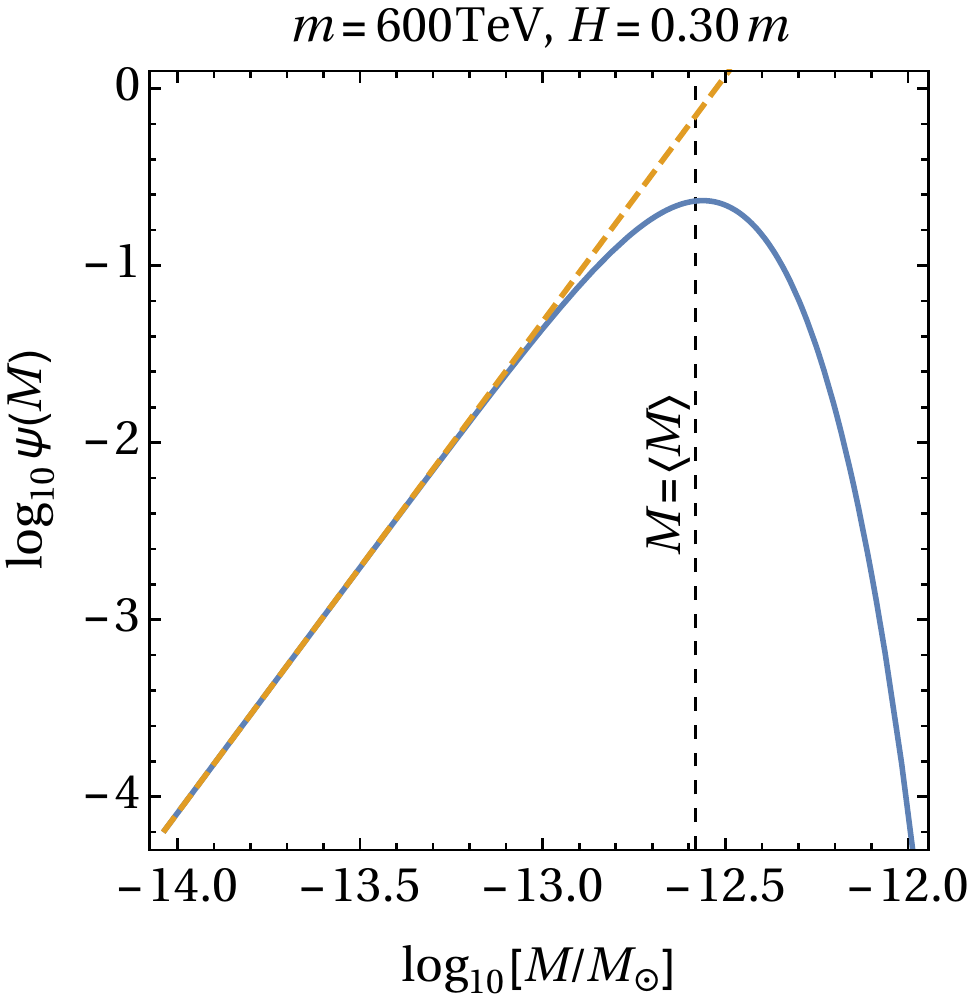} \hspace{10mm}
\includegraphics[height=0.36\textwidth]{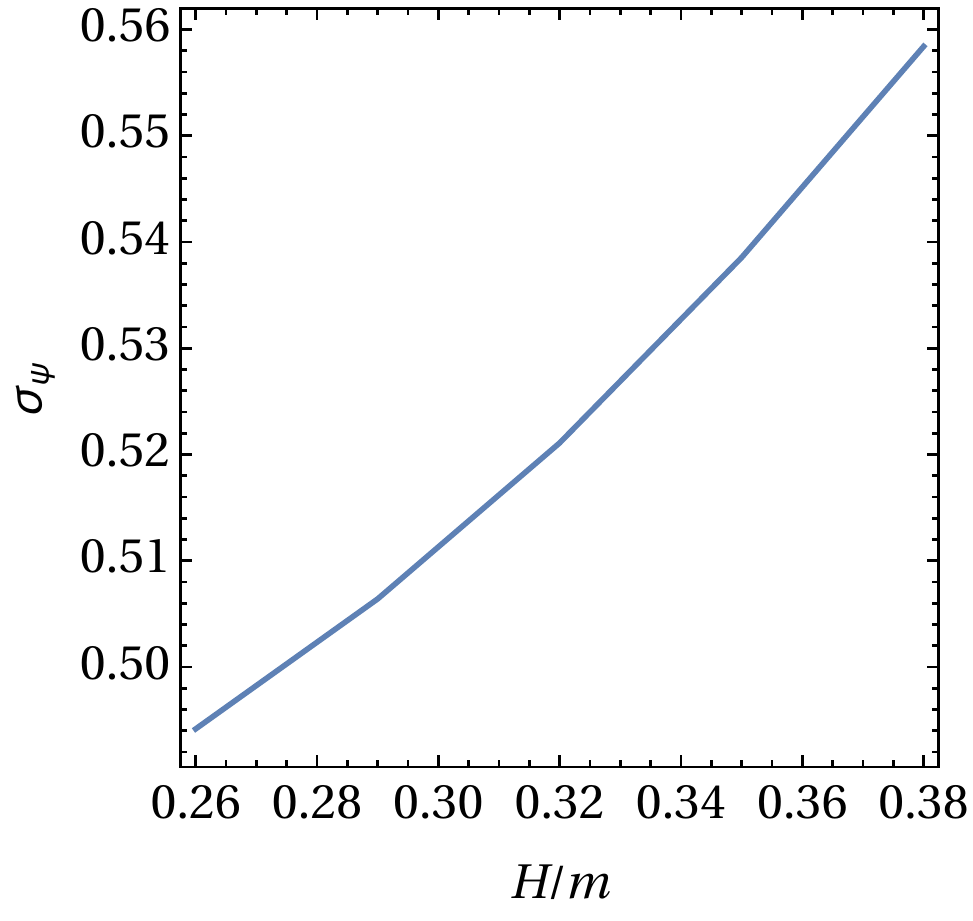}
\caption{\emph{Left panel:} The solid line shows the PBH mass spectrum produced from the perturbations formed during the late inflation for the parameters shown on top of the plot. The yellow dashed line shows power law behaviour, $\psi(M) \propto M^{1/\gamma}$, at small $M$. The expectation value of the distribution is indicated by the vertical dashed line. \emph{Right panel:} The width of the mass function as a function of $H/m$.}
\label{fig:mf}
\end{figure}

Next, let us calculate the PBH mass spectrum. When the scale $R$ reenters the horizon, the critical collapse in Eq.~\eqref{eq:Mcrit} gives rise to a mass spectrum~\cite{Niemeyer:1997mt,Yokoyama:1998xd}
\bea \label{eq:mf1}
\td \psi_R(M) &\equiv \td \Omega_R \int \td \delta\, P_R(\delta) \, \delta_D\left(\ln M_{{\rm BH},R}(\delta) - \ln M \right) = \td \Omega_R P_R(\delta) \frac{\td \delta}{\td \ln M} \\
&= \td \Omega_R \frac{q^{1/\gamma}}{\sqrt{2\pi} \sigma_R \gamma} \exp\left[-\frac{(\delta_c + q^{1/\gamma})^2}{2\sigma_R^2}\right]\,,
\eea
where $\delta_D$ denotes the Dirac delta function and $q\equiv M/(k M_R)$. Integrating over the comoving scales yields the total PBH mass spectrum today,
\be
\psi(M) = \int \td \psi_R(M) \,.
\ee
Notice that $\psi(M)$ is normalised to $\int \td \ln M \psi(M) = \Omega$. 

We have plotted an example of the resulting PBH mass function in the left panel of Fig.~\ref{fig:mf} for $\xi=1$. The shape of the mass function resembles the critical collapse mass spectrum~\eqref{eq:mf1}; It has a power law tail at small masses, $\psi(M)\propto M^{1/\gamma}$, and at large $M$ it drops exponentially. The expectation value of the mass distribution,
\be
\langle M \rangle = \int \td \ln M \, M\psi(M)\,,
\ee
shown by the vertical dashed line in the left panel of Fig.~\ref{fig:mf}, approximates well the peak mass of the spectrum. Moreover, the width of the mass spectrum (defined as in Ref.~\cite{Carr:2017jsz}),
\be
\sigma_\psi \equiv \sqrt{\langle \ln^2 M \rangle - \langle \ln M \rangle^2}
\ee
depends only on the ratio $H/m$, because that determines the shape of the curvature power spectrum. As shown in the right panel of Fig.~\ref{fig:mf}, the mass function is quite narrow.

\begin{figure}
\centering
\includegraphics[width=0.45\textwidth]{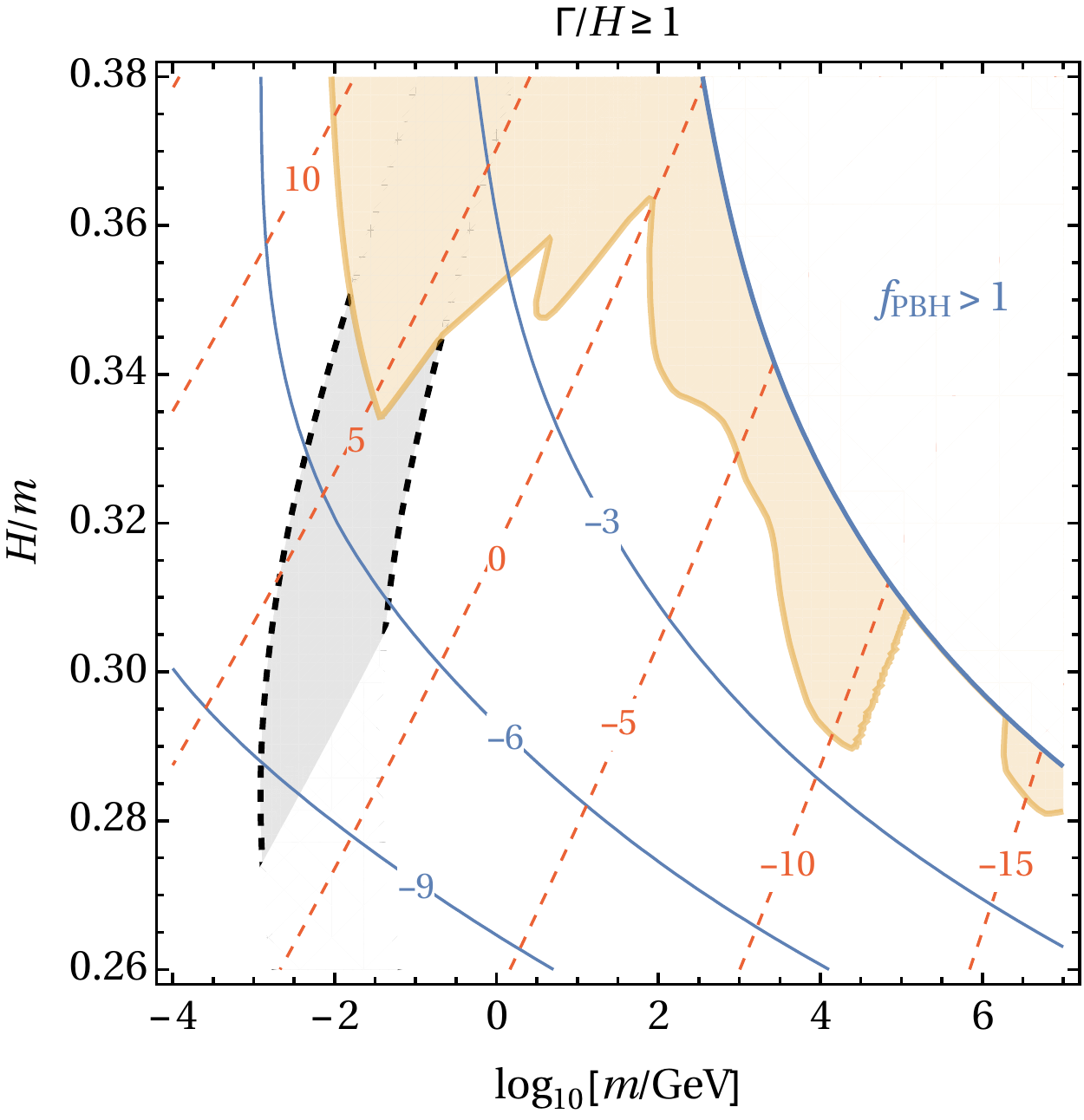} \hspace{4mm}
\includegraphics[width=0.45\textwidth]{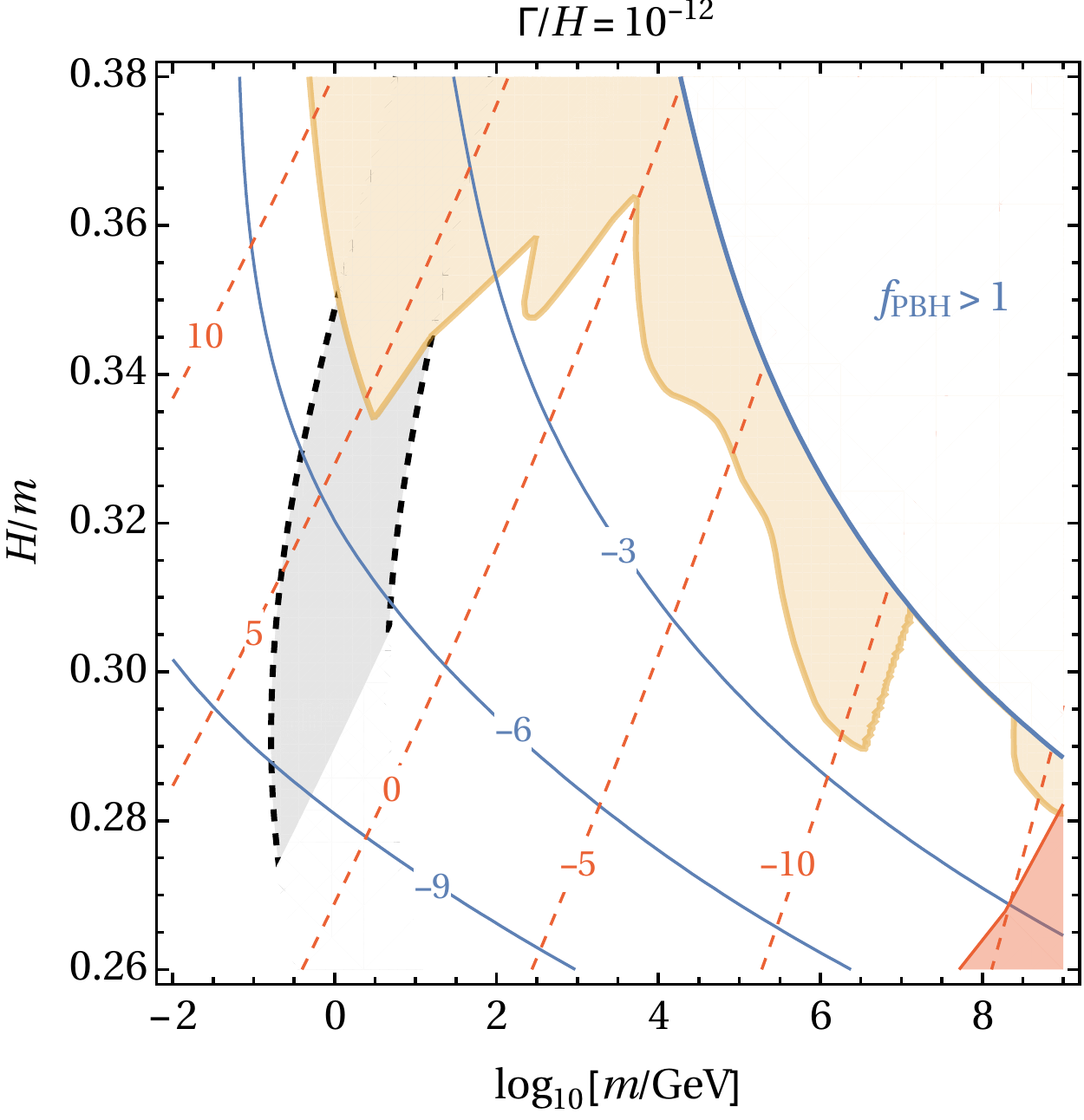}
\caption{The yellow curve shows the PBH constraints for monochromatic mass function (see e.g. Ref.~\cite{Carr:2016drx} for a review of the constraints), and the region between black dashed lines shows roughly the parameters that could possibly provide seeds for the SMBHs. The solid labeled contours indicate the fraction of dark matter in PBHs, $\log_{10} f_{\rm PBH}$, and the dashed ones the mean mass of the PBH mass spectrum, $\log_{10}[\langle M\rangle/M_\odot]$. Prompt reheating after late inflation has been assumed on the left panel, and a late reheating with $\Gamma/H = 10^{-12}$ on the right panel. In the shaded region in the lower right corner of the right panel the assumption that the scales relevant for PBH production re-enter horizon during radiation dominance is not valid.}
\label{fig:pbhs}
\end{figure}

The fraction of dark matter in PBHs, $f_{\rm PBH}\equiv \Omega/0.27$, and the mean mass of the produced spectrum, $\langle M \rangle$, are  shown in Fig.~\ref{fig:pbhs} as a function of the mass parameter $m$ and the Hubble rate during the inflation in units of $m$, $H/m$. In the upper right corner PBHs {are overproduced}. The solid blue lines indicate $\log_{10}f_{\rm PBH}$ and the red dashed ones $\log_{10}[\langle M\rangle/M_\odot]$. For the plot on the left panel we assumed prompt decay of $\phi$, but the result for a case that includes a period or matter dominance before $\phi$ decays can be obtained simply by rescaling the PBH abundance and mass as
\be
f_{\rm PBH} \propto \left(\frac{\Gamma}{H}\right)^\frac16\,, \quad \langle M \rangle \propto \left(\frac{\Gamma}{H}\right)^{-\frac13}\,.
\label{eq:G/H}
\ee
As an example, we show the case $\Gamma/H=10^{-12}$ on the right panel. The assumption that the scales relevant for PBH production re-enter horizon during radiation dominance is not valid in the red region.

Since the mass spectrum is very narrow, we can approximate the constraints on the PBH abundance simply by considering a monochromatic mass function with the same abundance at $M = \langle M \rangle$. This is shown by the yellow curve in Fig.~\ref{fig:pbhs}. The constraints combined for the yellow line arise from evaporation~\cite{Carr:2009jm}, femtolensing of gamma-ray bursts~\cite{Barnacka:2012bm}, microlensing results from Subaru/HSC~\cite{Niikura:2017zjd}, EROS~\cite{Tisserand:2006zx} and MACHO~\cite{Allsman:2000kg}, lensing of type Ia supernovae~\cite{Zumalacarregui:2017qqd}, LIGO observations~\cite{Wang:2016ana,Raidal:2017mfl,Ali-Haimoud:2017rtz,Raidal:2018bbj}, survival of a stars in dwarf galaxies~\cite{Brandt:2016aco,Koushiappas:2017chw}, and limits on accretion~\cite{Ricotti:2007au,Horowitz:2016lib,Ali-Haimoud:2016mbv,Poulin:2017bwe,Hektor:2018qqw}. We have discarded constraint claimed to arise from capture of PBHs by neutron stars~\cite{Capela:2013yf,Pani:2014rca}, the viability of which is questionable (see e.g. Ref.~\cite{Katz:2018zrn} for discussion). 
This opens up a window between $10^{-14}M_\odot$ and $10^{-11}M_\odot$ where all dark matter can be in PBHs. The parameters used for mass spectrum shown in the left panel of Fig.~\ref{fig:mf} are chosen such that the mass function is in this window, and the observed dark matter abundance is obtained in PBHs. As depicted in Fig.~\ref{fig:pbhs}, the generated PBHs can be the DM when \mbox{$m\simeq 600\,$TeV} and \mbox{$H/m\simeq 0.30$}.

Let us next consider the parameter space region that could provide seeds for SMBHs. We expect the PBH mass after formation to increase mainly due to accretion. At the Eddington limit, the mass of the PBH grows as~\cite{Salpeter:1964kb}
\be
M(t) = M_0 \exp\left(\frac{0.1}{\epsilon} \frac{t}{45\,{\rm Myr}}\right) \,,
\ee
where $\epsilon$ is the radiative efficiency, typically taken to be $\epsilon = 0.1$. In order to explain the observations of $\sim 10^9M_\odot$ BHs at redshifts $\sim 7$~\cite{Mortlock:2011va,Reed:2019ftq}, which is $760\,{\rm Myr}$ after the big bang, the seed BH mass has to be $M_0\gsim 50M_\odot$, with the lower bound corresponding to accretion at the Eddington limit with $\epsilon=0.1$ for $760\,{\rm Myr}$.\footnote{In the case of astrophysical seeds, the accretion time is much shorter, and therefore the seed BHs have to be much heavier, $M_0\gsim 10^3M_\odot$ for $\epsilon=0.1$.}

The seed PBH mass spectrum is very narrow, but we expect that it can become significantly wider due to differences in accretion histories, being possibly able to explain the whole SMBH mass spectrum that ranges form $10^5M_\odot$ to $10^9M_\odot$. To get a very rough estimate of the required seed PBH abundance, we vary the average SMBH mass between $10^5M_\odot$ and $10^9M_\odot$, and assume that they form $0.1\%$ of the dark matter density today~\cite{Merritt:2001zg}. The fraction of dark matter in PBHs before the accretion then has to be $10^{-3} M_{\rm ln}/(10^9M_\odot) \lsim f_{\rm PBH} \lsim - 10^{-3} M_{\rm ln}/(10^5M_\odot)$, and their mass $M_{\rm PBH} \gsim 50M_\odot$, for them to possibly provide the seeds for the SMBHs. This is indicated by the gray region between the black dashed lines in Fig.~\ref{fig:pbhs}, and requires $m\sim 0.01\,$GeV for prompt reheating. For late reheating, the plot is shifted according to Eq.~\eqref{eq:G/H}. For example, when 
\mbox{$\Gamma/H\sim 10^{-12}$} the region which provides the seeds for the SMBHs requires 
\mbox{$m\sim 1\,$GeV} as shown in Fig.~\ref{fig:pbhs}.

We conclude this section by pointing out that in our scenario for prompt reheating, $m\simeq 3\,{\rm GeV}$ and $H/m\simeq 0.35$ it is possible to produce large enough PBH abundance, $f_{\rm PBH}\sim 10^{-3}$~\cite{Raidal:2018bbj}, in the mass window where the LIGO events are. However, a careful analysis is needed to check how well the produced mass function fits the observations.

\section{Primordial inflation}
\label{sec:primeinf}

The presence of a late period of thermal inflation changes the predictions of primordial inflation, which generates the perturbations observed in the CMB. Primarily, this has to do with the substantial reduction of $N_*$, which is the number of remaining $e$-folds of primordial inflation after the time when the cosmological scales exit the horizon. In the usual case, when there is no subsequent inflation period after primordial inflation $N_*\simeq 50-60$ depending on the reheating efficiency after the primordial inflation. If reheating is prompt then $N_*\simeq 60$, while for late reheating it can decrease substantially down to 50 or so with \mbox{$T_{\rm reh}\sim 1\,$TeV}. However, in our scenario the total number of late inflation $e$-folds $N_T+N_R$ has to be subtracted from $N_*$, that is, $N_*=60(50)-N_T-N_R$ for prompt (late) primordial reheating. Because the predictions of primordial inflation, especially the scalar spectral index $n_s$ and the tensor-to-scalar ratio $r$, are determined by the value of $N_*$, we see that they are seriously affected in our case. The latest measurements indicate~\cite{Aghanim:2018eyx}
\be 
n_s = 0.965\pm 0.004 \quad \rm{and}\quad r<0.07 \,.
\label{eq:obs}
\ee

To evaluate the implications of our scenario we study the two possibilities of interest separately. 
Let us consider first the configuration that realises PBHs as dark matter candidates. Inserting the chosen values of $H/m\simeq 0.30$ and $m\simeq 600\,$TeV into Eq.~\eqref{eq:NR}, we get $N_R \simeq 15$. Then, using Eq.~\eqref{eq:NT} and $V_0 = 3m_P^2 H^2$ we get $N_T \simeq 13 + \ln g$. Therefore, the late inflation provides a total of $N_T+N_R \simeq 28 + \ln g$ $e$-folds and assuming prompt reheating for primordial inflation, this implies $N_* \simeq 32-\ln g$. Since we need $g>10^{-5}$ to have at least two $e$-folds of thermal inflation, the obtained $N_*$ is so small that the usual plateau inflation models, such as Starobinsky \cite{Starobinsky:1980te} or Higgs inflation~\cite{Bezrukov:2008ej}, are not appropriate. However, it has been recently demonstrated in Refs.~\cite{Dimopoulos:2016zhy,Dimopoulos:2016tzn} that thermal inflation may resurrect well motivated inflation models, which are ruled out otherwise because the spectrum of perturbations produced is not red enough. 

One such prominent example is the minimal hybrid inflation in supergravity~\cite{Linde:1997sj,Dvali:1997uq}. The model considers the most general superpotential with R-symmetry and a minimal K\"ahler potential. Respectively they are
\begin{equation}
W=\kappa S(\Phi\overline\Phi-M_{\rm GUT}^2)\quad{\rm and}\quad 
K=|\Phi|^2+|\overline\Phi|^2+|S|^2\,,
\end{equation}
with $\Phi$, $\overline\Phi$ being a pair of singlet lefthanded superfields and $S$ is a gauge singlet superfield which acts as the inflaton. The flatness of the inflationary trajectory is guaranteed by a U(1) R-symmetry on $S$. The parameter $M_{\rm GUT}\sim 10^{16}\,$GeV is the scale of a grand unified theory (GUT) (the VEV of the GUT Higgs field) and the parameter $\kappa\sim 0.1$ is a dimensionless coupling constant. Inflation proceeds along the $S$-direction, with the inflaton rolling down the Coleman-Weinberg one-loop radiative correction potential 
\begin{equation}
\Delta V\simeq\frac{\kappa^4M_{\rm GUT}^4}{16\pi^2}\,
\ln\left(\frac{\kappa^2|S|^2}{\Lambda^2}\right)\,, 
\label{eq:V:SUGRA}
\end{equation} 
where $\Lambda$ is some renormalisation scale. Inflation ends by triggering the GUT phase transition, when $|S|=M_{\rm GUT}$. The model employs the GUT-Higgs field as waterfall field,
while an accidental cancellation overcomes the $\eta$-problem without tuning. 

It was recently demonstrated that the model is resurrected when considering a period of late thermal inflation~\cite{Dimopoulos:2016tzn}. In this model, the spectral index and the tensor to scalar ratio are given by
\begin{equation} \label{nsrsugrahybrid}
n_s=1-\frac{1}{N_*}\quad{\rm and}\quad r=\frac{\kappa}{2\pi^2}\frac{1}{N_*}\,.
\end{equation}
As shown above, we have $N_*\simeq 32-\ln g$. Taking $g\sim 1$ we get $N_*\simeq 32$. Then, the above equation suggests $n_s\simeq 0.968$ and $r\sim 10^{-4}$, which are in good agreement with the observed values given in Eq.~\eqref{eq:obs}. The result is also quite sensitive to the primordial reheating. Indeed, if we consider the total number of $e$-folds (primordial plus late inflation) to be 57 instead of 60, then $N_*=29$, which results in
\mbox{$n_s=0.965$} that is the best fit value in Eq.~\eqref{eq:obs}.

Let us consider next the configuration that realises the seeds for the SMBHs. In this case we take $H/m\simeq 0.32$. As evident from Fig.~\ref{fig:pbhs}, the mass of the thermal inflaton field is smaller than the electroweak scale in this case. Therefore, in order for this field to remain unobserved today, its coupling $\hat g$ to the standard model particles must be suppressed. As we have discussed, when \mbox{$\hat g\ll 1$}, reheating after late inflation is not prompt. After the end of late inflation the Universe experiences a period of matter domination, until reheating finally takes place at temperature $T_{\rm reh}\sim\hat g\sqrt{m_P m}$, cf.~Eq.~\eqref{eq:Trehlate}. We consider the case when \mbox{$\hat g\sim 10^{-6}$}, which ensures that the thermal inflaton particle remains unobserved. Then, using that \mbox{$\Gamma\sim\hat g^2 m$}, we find \mbox{$\Gamma/H\sim 10^{-12}$}, in which case 
Fig.~\ref{fig:pbhs} in conjunction with Eq.~\eqref{eq:G/H} suggest \mbox{$m\simeq 1\,$GeV}. Repeating the same analysis as in the previous case,  we get $N_R \simeq 23$ and $N_T \simeq 20+\ln g$. Therefore, in this case the thermal plus fast-roll inflation provides a total of $N_T+N_R \simeq 43+\ln g$ $e$-folds. 
The total number of $e$-folds is reduced by the matter dominated period following the end of late inflation by \mbox{$\frac13\ln(V_0^{1/4}/T_{\rm reh})\simeq\frac16\ln(H/\Gamma)\simeq-\frac13\ln\hat g$} $e$-folds. Thus, in this case we have \mbox{$N_*\simeq 60+\frac13\ln\hat g-(43+\ln g)= 17-\frac13\ln(g^3/\hat g)$}. Using \mbox{$\hat g\sim 10^{-6}$}, we obtain \mbox{$N_*\simeq 12-\ln g$}. The reheating temperature is \mbox{$T_{\rm reh}\sim\hat g\sqrt{m_P m}\sim 1\,$TeV}, which means that reheating completes safely before BBN.
In view of Eq.~\eqref{nsrsugrahybrid}, we have $N_*=(1-n_s)^{-1}$. Thus, from the range in Eq.~\eqref{eq:obs}, to be in agreement with the observations, we require that $10^{-9}\lesssim g\lesssim 10^{-6}$. To make sure that our consideration of the Bunch-Davis vacuum is valid we need $N_T\simeq 20+\ln g\geq 2$, which means that the coupling range is truncated to $10^{-8}\lesssim g\lesssim 10^{-6}$. Taking for example $g\sim 10^{-7}$
we obtain $N_T\simeq 4$ and $n_s\simeq 0.965$, which is the best fit value in Eq. (\ref{eq:obs}).\footnote{It is straightforward to show that, assuming prolonged reheating after late inflation with \mbox{$\hat g\ll 1$}, can also improve the results in the dark matter case (\mbox{$m\simeq 600\,$TeV}), taking prompt primordial reheating.}
As with the previous case, \mbox{$r\sim 10^{-4}$}.

\section{An explicit model: Running mass thermal inflation}
\label{sec:runningmass}
In this section we introduce an explicit realisation for the unknown stabilising term of Eq.~\eqref{eq:pot} by replacing the constant mass term with a running one,
\begin{equation}
V(\phi,T)=V_0+\frac12m^2
\left[-1+\frac{\alpha}{32\pi^2}
  \ln\left(1+\frac{\phi^2}{\phi_0^2}\right)\right]\phi^2
+\frac12 g^2T^2\phi^2 \, ,
\label{Vrun}
\end{equation}
where $\alpha$ and $\phi_0$ are free parameters encoding information about the UV completion\footnote{The careful reader might argue that the same physics inducing the running mass in Eq.~\eqref{Vrun} would also generate a running quartic coupling for $\phi$. However, it can be shown that such term is absent in case of supersymmetric theories broken via soft terms \cite{Liddle:2000cg,Lyth:2009zz}. 
} that is inducing the running of the mass $m$. In case the running is generated by a boson of mass heavier than the thermal inflaton, we can estimate $\alpha$ and $\phi_0$ to be
\begin{equation} \label{phi0}
\phi_0 \sim \frac{M}{\sqrt h} \,, \quad
\alpha \sim h\left(\frac{M}{m}\right)^2  \,,
\end{equation}
where $h$ is a dimensionless coupling between the thermal inflaton and the heavier boson, while $M$ is its mass.

From the above we readily obtain the first and second derivative of the potential,
\begin{equation}
V'(\phi,T)=\frac{\alpha}{32\pi^2}m^2\frac{\phi^3}{\phi^2+\phi_0^2}+m^2
\left[-1+\frac{\alpha}{32\pi^2}
  \ln\left(1+\frac{\phi^2}{\phi_0^2}\right)\right]\phi
+g^2T^2\phi
\label{Vslope}
\end{equation}
and
\begin{equation}
V''(\phi,T)=
\frac{\alpha}{32\pi^2}m^2\phi^2\frac{3\phi^2+5\phi_0^2}{(\phi^2+\phi_0^2)^2}+m^2
\left[-1+\frac{\alpha}{32\pi^2}\ln\left(1+\frac{\phi^2}{\phi_0^2}\right)\right]
+g^2T^2\,.
\label{Vmass2}
\end{equation}
From these equations, we can easily check that effective mass squared at the origin is still given by Eq.~\eqref{eq:meff}. This means that thermal inflation proceeds exactly as discussed in Sec.~\ref{sec:Largepertubations}. Such a result was expected, because the argument in the logarithm is always bigger than unity, therefore the logarithmic term in Eq.~\eqref{Vrun} is always positive and serves to stabilise the potential, but it becomes important only when the field approaches its VEV, $\phi_{\rm vev}$. Thus, while the field is rolling down the potential hill, away from the origin, the logarithmic term is negligible and the potential reduces to the one in 
Eq.~\eqref{eq:pot}. This means that all the discussion  and results in the previous sections remain valid. In this section we will  investigate more about the new parameters of the model, $\alpha$ and $\phi_0$ (or alternatively $h$ and $M$). We will consider the two opposite cases\footnote{We are going to omit the case $\phi_{\rm vev}\sim\phi_0$ since it gives a similar phenomenology to the $\phi_{\rm vev}\ll\phi_0$ case.}: $\phi_{\rm vev}\ll\phi_0$ and $\phi_{\rm vev}\gg\phi_0$.

Let us start with the latter. Considering only the dominant terms for $\phi \gg \phi_0$, we can find that the VEV, defined by $V'(\phi_{\rm vev},T=0)=0$, can be approximated with
\begin{equation}
  \phi_{\rm vev} \simeq \phi_0
e^{\frac{16\pi^2}{\alpha}-\frac12}.
\label{phiVEV+}
\end{equation}
In a similar manner, a negligible cosmological constant, $V(\phi_{\rm vev},T=0)=0$, suggests
\begin{equation}
  V_0\simeq
  \frac{\alpha}{64\pi^2}m^2\phi_{\rm vev}^2\ll(m\phi_{\rm vev})^2 \, .
\label{V0run+}
\end{equation}
To have fast-roll inflation, we need $H\sim m$ which means
\begin{equation}
\frac{H^2}{m^2}\sim\frac{V_0}{m_P^2m^2}\sim
\frac{\alpha}{64\pi^2}\frac{\phi_{\rm vev}^2}{m_P^2}\sim 1
  \;\Rightarrow\;\phi_{\rm vev}\sim \frac{8\pi}{\sqrt\alpha}\,m_P\,,
\end{equation}
where we used Eq.~\eqref{V0run+}. Now, self-consistency of the used assumption $\phi_{\rm vev}\gg\phi_0$ implies $\sqrt\alpha\ll 4\pi$ and therefore a strongly super-Planckian  $\phi_{\rm vev}$. Thus, the potential in Eq.~\eqref{Vrun} cannot be trusted. Consequently, the case $\phi_{\rm vev}\gg\phi_0$ is excluded.

Let us now consider the first case. Using a Taylor expansion at the first order in $\phi/\phi_0 \ll 1$ and repeating the same steps as before, we find 
\begin{equation} \label{phiVEV}
\phi_{\rm vev} \simeq \frac{4\pi}{\sqrt\alpha}\,\phi_0\,, \quad
V_0 \simeq \frac14m^2\phi_{\rm vev}^2\,.
\end{equation}
Self-consistency of the used assumption $\phi_{\rm vev}\ll\phi_0$ now implies $\sqrt\alpha\gg 4\pi$. Finally, the vacuum mass, $m_V^2\equiv V''(\phi_{\rm vev}, T=0)$, is given by
\begin{equation}
m_V=\sqrt 2 m\,.
\label{mV}
\end{equation}
Moreover we notice that, by using $V_0\simeq 3H^2m_P^2$ and  Eq.~\eqref{phiVEV}, we obtain
\begin{equation}
\phi_{\rm vev} \simeq 2\sqrt3 m_P \frac{H}{m} \,.
\label{eq:vevnum}
\end{equation}
 Let us discuss now the two cases described in the previous section. We start with the the configuration that realises PBHs as dark matter candidates: $H/m\simeq 0.30$ and $m\simeq 600\,$TeV. We see immediately that $\phi_{\rm vev} \simeq 1.04 \, m_P$. As we have shown in Sec.~\ref{sec:primeinf}, inserting the chosen values of $H/m$ and $m$ into Eq.~\eqref{eq:NR}, we get $N_R \simeq 15$. Using Eqs.~\eqref{eq:NT} , \eqref{phiVEV} and~\eqref{eq:vevnum}, and taking \mbox{$g\sim 1$} we get $N_T \simeq 13$. Therefore the late inflation provides a total of $N_T+N_R \simeq 28$ e-folds. The running mass model contains two new free parameters, $h$ and $M$ (or alternatively $\alpha$ and $\phi_0$). Using Eq.~\eqref{phi0} and \eqref{phiVEV} we can derive
\be \label{eq:h}
h = \frac{4 \pi  m}{\phi_{\rm vev} } \simeq 12\,\frac{m}{m_P} \, .
\ee
In the present case we get\footnote{Such a coupling is quite small but not unrealistic, being, for instance, one order of magnitude bigger than the self-quartic coupling of an inflaton with a quartic potential.} $h\sim 10^{-12}$. For what concerns $M$, because of our expansion in $\phi_0/\phi_{\rm vev}\ll 1$, we cannot derive a numerical value but only a lower bound from the consistency condition $\sqrt\alpha\gg 4\pi$. Using Eqs.~\eqref{phi0},~\eqref{eq:vevnum} and \eqref{eq:h}, we get $M \gtrsim 2\sqrt{2\pi} V_0^{1/4}$.
 
We notice that the two independent energy scales $V_0$ and $M$ may turn out to have the same order of magnitude, leaving us the possible indication that they might have a common origin in the UV completion of the theory. To conclude we compute explicitly the scale  $V_0$ by using Eqs.~\eqref{phiVEV} and \eqref{eq:vevnum}, getting $V_0^{1/4} \sim 10^{12}\,$GeV, which is much larger than 1~MeV and therefore prompt reheating does not affect BBN. 

Let us consider now the configuration that realises seeds for the SMBHs: $H/m\simeq 0.32$ and $m\simeq 1\,$GeV. 
As discussed in Sec.~\ref{sec:primeinf}, in order to be in agreement with observations we need $g\ll 1$. Moreover, in this case the portal coupling becomes $h \sim 10^{-18}$ which is extremely small and not realistic. For completeness we provide also the values for the constant energy scale $V_0^{1/4}\sim 10^9\,$GeV.

\section{Conclusions}
\label{sec:conclusions}
We investigated the production of PBHs due to a tachyonic instability at the end of a period of thermal inflation, that is followed by fast-roll inflation given that the tachyonic mass of the thermal inflaton field is comparable to the Hubble scale of thermal inflation. We showed that the curvature power spectrum produced during this late inflation is peaked at scales that exit horizon $\mathcal{O}(1)$ $e$-folds after the onset of the phase transition. Significant fraction of the total energy density may then collapse gravitationally to form PBHs when these large fluctuations re-enter the horizon.

We found that this mechanism for PBH production can account for the dark matter in the Universe in its entirety at PBH masses $\sim 10^{-13}M_\odot$ if the tachyonic mass of our thermal inflaton field is $m\simeq 600\,$TeV, implying that also the mass of the thermal inflaton particle in the vacuum is of order $10^6\,$GeV. The thermal inflaton field can correspond to a flat direction in supersymmetry, in which case $m$ is simply a soft mass due to supersymmetry breaking. The supersymmetric setup also explains the absence of the quartic self-interaction terms since the contributions from bosons and fermions cancel out. 

The Hubble rate during the thermal (and fast-roll) inflation is in this case $\sim 180\,$TeV, meaning that the thermal inflation occurs well after primordial inflation. In total the late inflaton lasts for about 28 $e$-folds, thus the cosmological scales exit the horizon $N_*\simeq32$ $e$-folds before the end of primordial inflation. For such a small value of $N_*$ the usual plateau inflation models are not appropriate. Instead, we considered as an example the minimal hybrid inflation in supergravity that has been shown to benefit from the late inflationary period~\cite{Dimopoulos:2016tzn}. For $N_*\simeq 32$ it gives $n_s\simeq 0.968$, agreeing with the latest CMB observations. This can be improved if the reheating after primordial inflation or after late inflation is somewhat inefficient, bringing the result close to the best fit value of $n_s\simeq 0.965$.

In order to explain the SMBHs at galactic centres we need $m\simeq 1\,$GeV. Such a small value undermines any supersymmetric motivation in this case, and a particle with such mass would have been observed in LHC, unless its coupling with the standard model particles is very small. This implies that the Universe is reheated well after the end of fast-roll inflation. Taking $\hat g\sim 10^{-6}$ and $g\sim 10^{-7}$ suggests that the cosmological scales exit the horizon about $N_*\simeq 28$ $e$-folds before the end of primordial inflation. Considering again minimal hybrid inflation in supergravity we obtain the spectral index $n_s\simeq 0.965$, which is in excellent agreement with the observations.

Although our treatment of PBH formation is generic, we have also constructed an explicit model realisation for the thermal inflaton field, whose potential is stabilised by the loop correction to its mass. Such construction introduces two new free parameters: $h$, a dimensionless coupling between the thermal inflaton and a heavier boson, and $M$, its mass. We obtain a reasonable region of the parameters space for the PBH DM, $h \sim 10^{-12}$ and $M \gtrsim 10^{12}\,$GeV, very close to the typical order of magnitude of the parameters of a primordial inflaton, and a possible direction about the UV completion of the model.  On the contrary, explaining the SMBH at galactic centres requires the unrealistic value $h \sim 10^{-18}$, meaning that, in this case, the stabilisation of the thermal inflaton potential is due to another mechanism rather than its running mass.

\section*{Acknowledgements}
We thank Bernard Carr, Marek Lewicki and Tommi Tenkanen for useful discussions. KD is supported by the Lancaster-Manchester-Sheffield Consortium for Fundamental Physics under the STFC grant ST/L000520/1, TM by the STFC grant ST/P000762/1 and by the Estonian Research Council via the Mobilitas Plus grant MOBJD323, VV by the STFC Grant ST/P000258/1 and AR is supported by the Estonian Research Council grant PUT1026, the grant IUT23-6 of the Estonian Ministry of Education and Research, and by the EU through the ERDF CoE program project TK133.

\bibliography{MBHs.bib}

\end{document}